\documentclass[conference]{IEEEtran}
\IEEEoverridecommandlockouts
\usepackage{cite}
\usepackage{amsmath,amssymb,amsfonts}
\usepackage{amsthm}
\usepackage{algorithm}
\usepackage{algcompatible}
\usepackage{graphicx}
\usepackage{bm}
\usepackage{textcomp}
\usepackage{xcolor}

\def\BibTeX{{\rm B\kern-.05em{\sc i\kern-.025em b}\kern-.08em
    T\kern-.1667em\lower.7ex\hbox{E}\kern-.125emX}}

\begin{document}
\title{Error Rate Analysis for Grant-free Massive Random Access with Short-Packet Transmission\\
}

\author{\IEEEauthorblockN{Xinyu Bian$^{\ast}$, Yuyi Mao$^{\dagger}$, and Jun Zhang$^{\ast}$}
\IEEEauthorblockA{{$^{\ast}$Dept. of ECE, The Hong Kong University of Science and Technology, Hong Kong}\\
{$^{\dagger}$Dept. of EIE, The Hong Kong Polytechnic University, Hong Kong} \\
Email: xinyu.bian@connect.ust.hk, yuyi-eie.mao@polyu.edu.hk, eejzhang@ust.hk}
}

\maketitle
\begin{abstract}
Grant-free massive random access (RA) is a promising protocol to support the massive machine-type communications (mMTC) scenario in 5G and beyond networks. In this paper, we focus on the error rate analysis in grant-free massive RA, which is critical for practical deployment but has not been well studied. We consider a two-phase frame structure, with a pilot transmission phase for activity detection and channel estimation, followed by a data transmission phase with coded data symbols. Considering the characteristics of short-packet transmission, we analyze the block error rate (BLER) in the finite blocklength regime to characterize the data transmission performance. The analysis involves characterizing the activity detection and channel estimation errors as well as applying the random matrix theory (RMT) to analyze the distribution of the post-processing signal-to-noise ratio (SNR). As a case study, the derived BLER expression is further simplified to optimize the pilot length. Simulation results verify our analysis and demonstrate its effectiveness in pilot length optimization. 
\end{abstract}
  
\begin{IEEEkeywords}
Grant-free massive random access, approximate message passing (AMP), short-packet transmission, block error rate (BLER), random matrix theory (RMT), pilot length optimization.
\end{IEEEkeywords}

\section{Introduction}
Massive machine-type communications (mMTC) is an important application scenario in the fifth generation (5G) wireless networks with the exploding number of connected devices \cite{b1}. Although massive devices are expected to transmit short data packets in mMTC, only a small fraction of them are active in a given time slot, and novel access schemes are required for this scenario \cite{b2}. As a result, various grant-free random access (RA) schemes have been proposed, where each active user can directly transmit the payload data without a grant from the base station (BS), to avoid the long access latency and huge signalling overhead \cite{b3}.

A two-phase transmission protocol is typically adopted for grant-free massive RA, where the pilot symbols are first transmitted to detect the set of active users and estimate their channel coefficients, followed by the data transmission phase to delivery the payload \cite{b4}. However, due to the limited radio resources reserved for pilot transmission, only non-orthogonal pilot sequences can be assigned to the vast amount of users, which poses great challenges for accurate user activity detection and channel estimation. Fortunately, thanks to the sporadic traffic pattern of mobile and Internet of Things (IoT) devices, user activity detection and channel estimation for grant-free massive RA can be viewed as a compressive sensing problem \cite{b5}, for which, many efficient algorithms have been developed.

Specifically, by applying the orthogonal matching pursuit (OMP) algorithm, joint activity and data detection was performed for non-orthogonal multiple access systems in \cite{b6}. An approximate message passing (AMP) method with expectation maximization (EM) was proposed to tackle a similar problem in \cite{b7}. Although these methods exhibit great potential in detecting the user activity, they assume perfect channel state information (CSI) that is unrealistic for practical systems. Therefore, joint activity detection and channel estimation (JADCE) has received intensive recent attention \cite{b8,b9,b10,b11}. In \cite{b8}, an AMP algorithm with a minimum mean square error (MMSE) denoiser was proposed for JADCE. To improve the accuracy of JADCE, channel sparsity from both the spatial and angular domains was also utilized \cite{b9}. Besides, a structured group sparsity estimation method was developed in \cite{b10}, and the optimal pilot length is determined according to the phase transition behavior. In addition, the common sparsity in pilot and data signal, as well as the soft information from a channel decoder, were exploited in \cite{b11} for joint activity detection, channel estimation, and data decoding.

The performance of grant-free massive RA systems has also been analyzed in a few prior studies. For example, the achievable rate of grant-free massive RA systems was characterized in \cite{b12} assuming perfect user activity detection, but the inﬁnite blocklength coding may not be suitable for mMTC with short-packet transmission. Hence, the successful symbol transmission rate (SSTR) was proposed as a more reasonable performance metric for grant-free massive RA in \cite{b13}. However, the analysis was limited to uncoded transmission without considering the user activity detection error. As a result, it necessitates more comprehensive data transmission performance analysis for grant-free massive RA.

In this paper, we analyze the data error rate for grant-free massive RA by considering the classical two-phase frame structure. By adopting the AMP algorithm for JADCE, the user activity detection errors, including the false alarm and missed detection probabilities, are characterized based on the state evolution. Conditioned on the outcome of user activity detection, the block error rate (BLER) is obtained by analyzing the distribution of the receive signal-to-noise ratio (SNR) via the random matrix theory (RMT). Then, we simplify our analysis by deriving a tight approximation in close-form, which facilitates efficient optimization of the pilot length. Simulation results corroborate our theoretical analysis and show the significance of pilot length optimization for grant-free massive RA systems.

The rest of this paper is organized as follows. We introduce the system model and the corresponding communication protocol in Section \uppercase\expandafter{\romannumeral2}. The data error rate is analyzed in Section \uppercase\expandafter{\romannumeral3}, and pilot length optimization for grant-free massive RA is investigated in Section \uppercase\expandafter{\romannumeral4}. Simulation results are presented in Section \uppercase\expandafter{\romannumeral5}, and conclusions are drawn in Section \uppercase\expandafter{\romannumeral6}.

\textbf{Notations:} We use lower-case letters, bold-face lower-case letters, bold-face upper-case letters, and math calligraphy letters to denote scalars, vectors, matrices, and sets, respectively. The transpose and conjugate transpose of a matrix $\mathbf{M}$ are denoted as $\mathbf{M}^{\mathrm{T}}$ and $\mathbf{M}^{H}$, respectively. Besides, we denote the complex Gaussian distribution with mean $\bm{\mu}$ and covariance matrix $\bm{\Sigma}$ as $\mathcal{C} \mathcal{N}(\bm{\mu}, \bm{\Sigma})$, and the Dirac delta function as $\delta_{0}$. In addition, $\mathbb{E}[\cdot]$ and $\operatorname{Var}[\cdot]$ denote the statistical expectation and variance, respectively.

\section{System Model and Communication Protocol}
\begin{figure}[htpb]
\centering
\includegraphics[width=3in]{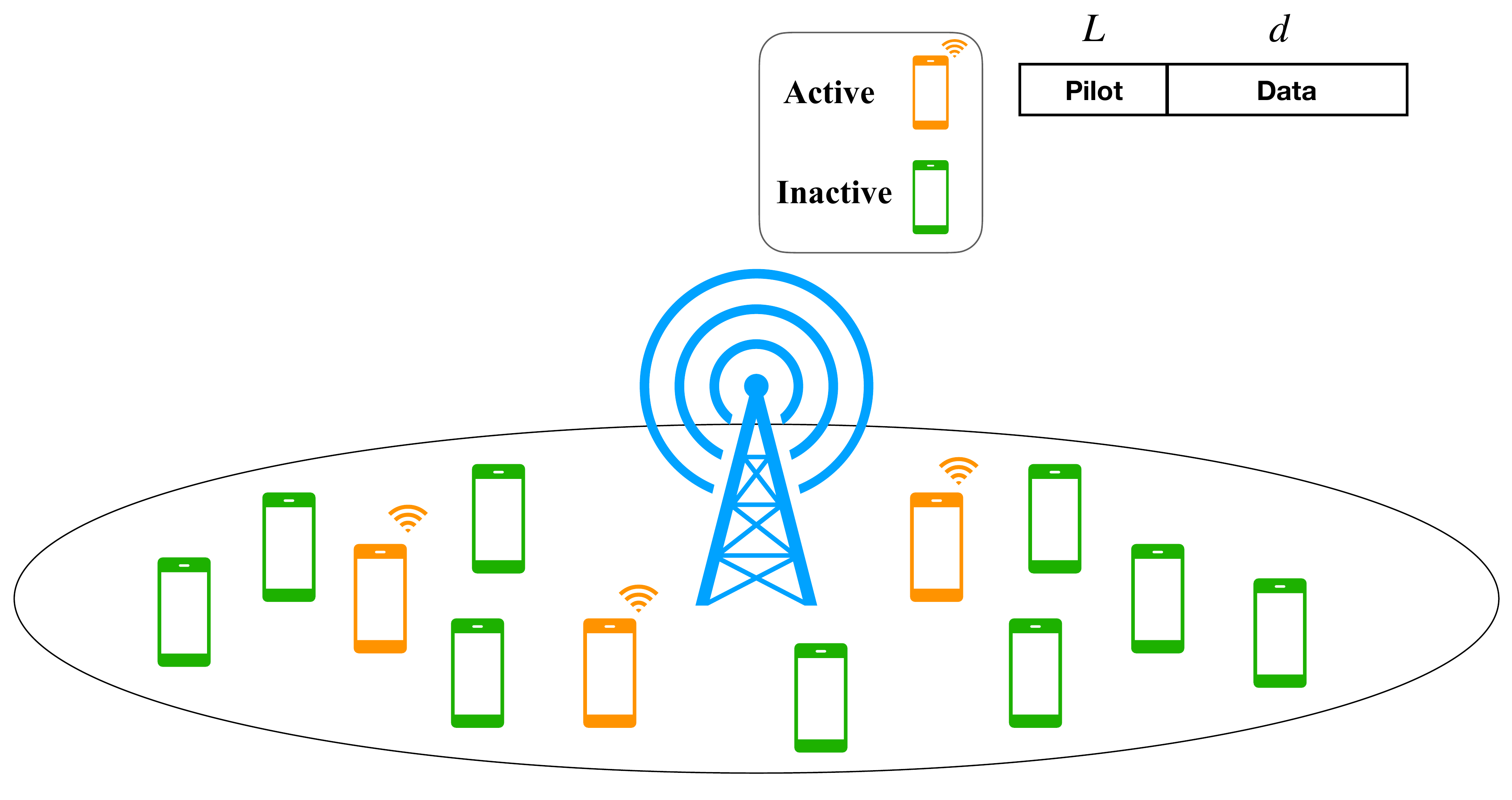}
\caption{System model and the frame structure for grant-free massive RA.}
\label{model}
\end{figure}

We consider an uplink cellular network that consists of $N$ single-antenna users and an $M$-antenna BS, as shown in Fig. \ref{model}. The set of users is denoted as $\mathcal{N}\triangleq \{1,\cdots,N\}$. It is assumed that $K$ out of the $N$ users are active for transmission at each time, and all the users become active with probability $\lambda$. We use $u_{n} \in \{0,1\}$ to denote the user activity, where $u_{n}=1$ indicates user $n$ is active and vice versa. Thus, the set of active users can be represented by $\mathcal{K} \triangleq \left\{n \in \mathcal{N} | u_{n}=1 \right\}$. We further assume the number of BS antennas is no less than the number of active users, i.e.,  $M\geq K$, to avoid the system from being overloaded \cite{b14}.

Each transmission block spans $T$ symbol intervals, and the block fading channel model is adopted, i.e., the channel condition remains unchanged within a transmission block. The uplink channel vector from user $n$ to the BS is denoted as $\tilde{\mathbf{h}}_{n}=\sqrt{\beta_{n}} \boldsymbol{\alpha}_{n}$, where $\boldsymbol{\alpha}_{n}$ and $\beta_{n}$ represent the small-scale and large-scale fading coefficients, respectively, and $\{\beta_n\}$'s are assumed to be known at the BS. The classical two-phase grant-free RA scheme \cite{b9} is employed, where in each transmission block, the first $L$ symbols are reserved for pilot transmission and the remaining $d\triangleq T-L$ symbols are used for data delivery.

Since $N$ is much greater than $L$ in massive RA systems, it is infeasible to assign orthogonal pilots to all the users. Instead, we assign each user with a unique pilot sequence, denoted as $\sqrt{L}\mathbf{p}_{n}$ with $\mathbf{p}_{n}\triangleq \left[p_{n, 1}, \cdots, p_{n, L}\right]^{\mathrm{T}}$ and $p_{n, l} \sim \mathcal{C} \mathcal{N}\left(0, \frac{1}{L}\right)$, which are asymptotic orthogonal when $L\rightarrow \infty$ \cite{b8}. To compensate the path loss, we implement the statistical channel inversion power control \cite{b15}, such that the average received signal strength at the BS meets a target value $\beta$. Then, the received pilot signal $\mathbf{Y}_{p} \in \mathbb{C}^{L\times M}$ at the BS can be expressed as follows: 
\begin{align}
\mathbf{Y}_{p}=\sum_{n \in \mathcal{N}}\sqrt{L}u_{n}\mathbf{p}_{n}\mathbf{h}^{\mathrm{T}}_{n}+\mathbf{N}_{p},
\end{align}
\noindent where $\mathbf{h}_{n} \triangleq \sqrt{\beta}\boldsymbol{\alpha}_{n}$, and $\mathbf{N}_{p}=\left[\mathbf{n}_{p,1},...,\mathbf{n}_{p,L}\right]^{\mathrm{T}}$ denotes the Gaussian noise with zero mean and variance $\sigma^{2}$ for each element. We adopt the AMP algorithm developed in \cite{b9} to perform joint activity detection and channel estimation, considering both its superior performance and rich theoretical understanding \cite{b12}. In particular, based on the received pilot signal $\mathbf{Y}_{p}$, the AMP algorithm is able to obtain an estimate of $\{\mathbf{h}_{n}\}$, denoted as $\{\hat{\mathbf{h}}_{n}\}$, and a set of users $\hat{\mathcal{K}}=\{j \in \mathcal{N}| \hat{u}_{j}=1\}$ that are detected as active, where $\hat{u}_{j}$ is an indicator of the estimated user activity. We further denote $\Delta \mathbf{h}_{n}\triangleq\mathbf{h}_{n}-\hat{\mathbf{h}}_{n}$ as the channel estimation error of user $n$.

In the data transmission phase, each active user transmits a packet containing $c$ information bits, which are mapped to a codeword of $d$ symbols $\mathbf{s}_{n}$ by a channel encoder. Thus, the channel coding rate is given as $R=c/d$. Assuming unit power of each symbol, the received data signal at the BS $\mathbf{Y}\in \mathbb{C}^{M \times d}$ can be expressed as follows: 
\begin{align}
\mathbf{Y}= \sum_{n \in \mathcal{K}} \mathbf{h}_{n} \mathbf{s}_{n}^{\mathrm{T}}+\mathbf{N},
\end{align}

\noindent where the noise term $\mathbf{N}=\left[\mathbf{n}_{1},...,\mathbf{n}_{d}\right]$ follows the same distribution as $\mathbf{N}_{p}$.

To decode the payload data, a low-complexity zero-forcing (ZF) equalizer is first applied to eliminate the inter-user interference. As a result, the post-processing data symbols of an active user that is correctly detected can be written as follows:
\begin{align}
\begin{aligned}
\hat{\mathbf{s}}_{k}&=\hat{\mathbf{w}}_{k}^{H}\mathbf{Y}= \hat{\mathbf{w}}_{k}^{H} \left(\sum_{n \in \mathcal{K}} \mathbf{h}_{n} \mathbf{s}_{n}^{\mathrm{T}}+\mathbf{N}\right)\\
&= \hat{\mathbf{w}}_{k}^{H} \left(\sum_{n \in \mathcal{K}} \left(\hat{\mathbf{h}}_{n}+\Delta \mathbf{h}_{n}\right) \mathbf{s}_{n}^{\mathrm{T}}+\mathbf{N}\right)\\
&=\underbrace{\hat{\mathbf{w}}_{k}^{H} \hat{\mathbf{h}}_{k}\mathbf{s}_{k}^{\mathrm{T}}}_{\text{Signal}}+\underbrace{\hat{\mathbf{w}}_{k}^{H}\sum_{n \in \mathcal{K},n \neq k} \hat{\mathbf{h}}_{n}\mathbf{s}_{n}^{\mathrm{T}}}_{\text{Inter-user interference}}\\
&\underbrace{+\hat{\mathbf{w}}_{k}^{H}\sum_{n \in \mathcal{K}} \Delta \mathbf{h}_{n}\mathbf{s}_{n}^{\mathrm{T}}}_{\text{Channel\ estimation\ error}}+\underbrace{\hat{\mathbf{w}}_{k}^{H}\mathbf{N}, k\in\hat{\mathcal{K}}}_{\text{Noise}},
\end{aligned}
\end{align}

\noindent where $\hat{\mathbf{w}}_{k}$ is the $k$-th column of the following ZF equalizer:
\begin{align}
\hat{\mathbf{W}}= \hat{\mathbf{H}}\left(\hat{\mathbf{H}}^{H}\hat{\mathbf{H}}\right)^{-1},
\end{align}
\addtolength{\topmargin}{0.04in}
\noindent and $\hat{\mathbf{H}}\triangleq \left[\{ {\hat{\mathbf{h}}_{j}}\}_{ j \in \hat{\mathcal{K}}} \right]$. Since the ZF equalizer can effectively mitigate inter-user interference only when $M\geq |\hat{\mathcal{K}}|$, we assume no data symbol can be correctly detected when $M<|\hat{\mathcal{K}}|$, and in this case transmission errors occur for all the active users. This can be regarded as a conservative estimation of the system performance \cite{b16}. 

In order to analyze the data error rate, it is critical to obtain the post-processing SNR distribution according to (3). However, since the set of active users are unknown at the BS, the post-processing SNR in grant-free massive RA systems also depends on the outcome of user activity detection. The coupling of user activity detection and data detection makes the analysis non-trivial.

\section{Data Error Rate Analysis}
In this section, we carry out the data error rate analysis for grant-free massive RA systems. Note that as no data detection and decoding is performed for the missed detection users, the BLER of an active user can be expressed as follows:
\begin{align}
    P_{e,k}&=P_{M,k}+\left(1-P_{M,k}\right)\varepsilon_{k}, k\in\mathcal{K},
\end{align}

\noindent where $P_{M,k}\triangleq P(\hat{u}_{k}=0|u_{k}=1)$ denotes the missed detection probability of user $k$, and $\varepsilon_{k}$ is the average BLER of user $k$ given it is correctly detected. To evaluate (5), we first review the theoretical results of the AMP algorithm for JADCE, and introduce the error rate analysis of finite blocklength channel coding.

\subsection{Preliminaries}
\subsubsection{AMP for JADCE}
Based on the received pilot signal, the AMP algorithm \cite{b8} iteratively recovers the effective channel coefficients $\{u_{n}\mathbf{h}_{n}\}$ for all users with their sparse priors. When the algorithm converges, the set of active users is determined from the estimated effective channel coefficients by a thresholding operation. The performance of the AMP algorithm is measured by the mean square error (MSE), which can be accurately predicted by the state evolution as elaborated in Lemma 1 that forms the basis of our analysis. In this paper, we focus on the regime with $L>K$, where the AMP algorithm was reported to have a stable behavior \cite{b12}.

\textbf{Lemma 1} [8, Proposition 9].
The state evolution of the AMP algorithm can track the MSE of the estimated effective channel coefficients via the following recursion:
\begin{align}
\begin{aligned}
\mathbf{\Sigma}_{t+1}=\frac{\sigma^{2}}{L} \mathbf{I}+\frac{N}{L} \mathbb{E}_{\mathbf{X},\mathbf{V}}\left[\left(\eta_{t}\left(\mathbf{X}+\mathbf{\Sigma}_{t}^{\frac{1}{2}} \mathbf{V}\right)-\mathbf{X}\right)\right. \\
\left.\left(\eta_{t}\left(\mathbf{X}+\mathbf{\Sigma}_{t}^{\frac{1}{2}} \mathbf{V}\right)-\mathbf{X}\right)^{H}\right],
\end{aligned}
\end{align}

\noindent where $\mathbf{\Sigma}_{t}=\tau_{t}^{2}\mathbf{I}$ is the state in the $t$-th iteration, and $\eta_{t}(\mathbf{X}+\mathbf{\Sigma}_{t}^{\frac{1}{2}} \mathbf{V})=\mathbb{E}[\mathbf{X} \mid \mathbf{X}+\mathbf{\Sigma}_{t}^{\frac{1}{2}} \mathbf{V}]$ denotes the MMSE denoiser. Besides, $\mathbf{X}$ and $\mathbf{V}$ are random variables with $\mathbf{X} \sim (1-\lambda) \delta_{0}+\lambda \mathcal{CN}(0,\beta\mathbf{I})$ and $\mathbf{V} \sim \mathcal{CN}(0,\mathbf{I})$.

When the AMP algorithm converges, the state converges to a fixed point $\tau_{\infty}$ that can be used to derive the probabilities of missed detection and false alarm for user activity detection as follows:
\begin{align}
P_{M,n}&= P_{M} \triangleq \frac{1}{\Gamma(M)} \bar{\gamma}\left(M, l^{2}\left(\beta^{2}+\tau_{\infty}^{2}\right)^{-1}\right), n\in\mathcal{K},\\
P_{F,n}&= P_{F} \triangleq 1-\frac{1}{\Gamma(M)} \bar{\gamma}\left(M, l^{2} \tau_{\infty}^{-2}\right), n\in \mathcal{N}\setminus \mathcal{K},
\end{align}

\noindent where $\Gamma(\cdot)$ and $\bar{\gamma}(\cdot)$ denote the Gamma function and the upper incomplete Gamma function, respectively, and $l\triangleq M \ln \left(1+\frac{\beta}{\tau_{\infty}^{2}}\right) /\left(\frac{1}{\tau_{\infty}^{2}}-\frac{1}{\tau_{\infty}^{2}+\beta}\right)$. Besides, according to Theorem 3 in \cite{b12}, the channel estimation error $\Delta \mathbf{h}_{k}$ of an active user is zero mean and the covariance matrix is given as follows:
\begin{align}
\operatorname{Cov}\left(\Delta \mathbf{h}_{k}, \Delta \mathbf{h}_{k}\right)=\Delta v_{k} \mathbf{I}, k \in \mathcal{K},
\end{align}

\noindent where $\Delta v_{k}=\frac{\beta \tau_{\infty}^{2}}{\beta+\tau_{\infty}^{2}}$ for $k \in \mathcal{K}\cap \hat{\mathcal{K}}$ and $\Delta v_{k}=\beta$ for $k \in \mathcal{K} \setminus \hat{\mathcal{K}}$.

\subsubsection{BLER of finite-blocklength channel coding}
Short-packet transmission is a key technology to support the new services in mMTC. However, the Shannon capacity formula assumes asymptotic blocklength and thus cannot provide an accurate performance characterization in the finite blocklength regime. To bridge this gap, Polyanskiy \emph{et al.} shows in \cite{b17} that the maximum achievable rate in the finite blocklength regime can be tightly approximated by subtracting a backoff term from the channel capacity, as detailed in the following lemma.

\textbf{Lemma 2} [17].
The maximum achievable rate of an additive white Gaussian noise (AWGN) channel in the finite blocklength regime can be tightly approximated as follows:
\begin{align}
    R(d,\varepsilon)\approx C(\gamma)-\sqrt{\frac{V(\gamma)}{d}}Q^{-1}(\varepsilon),
\end{align}

\noindent where $d$ is the blocklength, $\varepsilon$ is the error probability, and $\gamma$ is the SNR. The term $C(\gamma)=\log_{2}{(1+\gamma)}$ is the channel capacity and $V(\gamma)=\frac{\gamma(\gamma+2)}{2(\gamma+1)^2}\log_{2}^2{e}$ is the channel dispersion. Besides, $Q^{-1}(x)$ denotes the inverse function of $Q(x)\triangleq \int_{x}^{\infty} \frac{1}{\sqrt{2 \pi}} e^{-t^{2} / 2} dt$.

Accordingly, the BLER of an active user that is correctly detected can be approximated as follows:
\begin{align}
   \varepsilon_{k}\left(\gamma_{k}\right) \approx Q\left(\frac{C(\gamma_{k})-R}{\sqrt{V(\gamma_{k})/d}}\right), k \in \mathcal{K}\cap \hat{\mathcal{K}},
\end{align}

\noindent where $\gamma_{k}$ is the post-processing SNR as will be analyzed in the next subsection.

\subsection{Post-processing SNR and BLER Analysis}
Based on the signal model in (3), the post-processing SNR of an active user that is successfully detected can be expressed as follows:
\begin{align}
\begin{aligned}
    \gamma_{k}&=\mathbb{E}\left[|| \hat{\mathbf{w}}_{k}^{H} \hat{\mathbf{h}}_{k}\mathbf{s}_{k}^{T}||^2\right]\Big/\mathbb{E}\Big[\Big(||\hat{\mathbf{w}}_{k}^{H}\sum_{n \in \mathcal{K},n \neq k}  \hat{\mathbf{h}}_{n}\mathbf{s}_{n}^{T}||^2\\
    &+||\hat{\mathbf{w}}_{k}^{H}\sum_{n \in \mathcal{K}}  \Delta \mathbf{h}_{n}\mathbf{s}_{n}^{T}||^2+||\hat{\mathbf{w}}_{k}^{H}\mathbf{N}||^2\Big)\Big]\\
    &\overset{(\text{a})}{=}\frac{d}{0+||\hat{\mathbf{w}}_{k}^{H}||^2\left(\sum_{n \in \mathcal{K}} \mathbb{E}\left[||\Delta \mathbf{h}_{n}||^2\right]d+\mathbb{E}\left[||\mathbf{N}||^2\right]\right)}\\
    &\overset{(\text{b})}{=}\frac{1}{\left[\left(\hat{\mathbf{H}}^{H}\hat{\mathbf{H}}\right)^{-1}\right]_{kk} \left(\sum_{n \in \mathcal{K}} \operatorname{Var}(\Delta \mathbf{h}_{n})+\sigma^2\right)},
\end{aligned}   
\end{align}

\noindent where (a) is because $\mathbb{E}\left[\mathbf{s}_{n}^{T}\mathbf{s}_{n}\right]=d$ and $\hat{\mathbf{H}}^{H}\hat{\mathbf{W}}=\mathbf{I}$, and (b) is derived using $\Delta \mathbf{h}_{n} \sim \mathcal{CN}(\mathbf{0},\Delta v_{n}\mathbf{I})$, $ n\in\mathcal{K}$. 

It is observed from (12) that the distribution of $\gamma_{k}$ depends on both the missed detection and false alarm users. Specifically, the term $\sum_{n \in \mathcal{K}}\operatorname{Var}(\Delta \mathbf{h}_{n})$ contains all the active users with different channel estimation errors depending on whether they are detected correctly. Besides, the dimension of matrix $\hat{\mathbf{H}}^{H}\hat{\mathbf{H}}$ equals the number of users that are estimated as active, which influences the distribution of $[(\hat{\mathbf{H}}^{H}\hat{\mathbf{H}})^{-1}]_{kk}$. Therefore, it is important to consider different user activity detection outcomes when analyzing the post-processing SNR distribution, as elaborated in the following lemma.

\textbf{Lemma 3}.
Suppose there are $e$ ($e=0,1,\cdots,K$) missed detection and $f$ ($f=0,1,\cdots,N-K$) false alarm users. If $M\geq|\hat{\mathcal{K}}|= K-e+f$, the post-processing SNR of a correctly detected active user can be rewritten as follows:
\begin{align}
    \gamma_{k|ef}\!=\!\frac{1}{\left[\left(\hat{\mathbf{H}}^{H}\hat{\mathbf{H}}\right)^{-1}\right]_{kk}\!\left((K-e)\frac{\beta\tau_{\infty}^2}{\beta+\tau_{\infty}^2}+e \beta+\sigma^2\right)},
\end{align}


\noindent which is a Chi-square random variable with the probability density function (PDF) given as $f_{\gamma_{k|ef}}(x)=\frac{x^{\theta_{2}-1}e^{-x/(2\theta_1)}}{2^{\theta_2}\Gamma(\theta_2)}$ $\left(x\geq 0\right)$, and $\theta_1=\frac{\beta^2}{(K-e)\beta\tau_{\infty}^2+(e \beta+\sigma^2)(\beta+\tau_{\infty}^2)}$ and $\theta_2=M-K+e-f+1$. 

\begin{proof}
Since the columns in $\hat{\mathbf{H}}$ are linearly independent, according to the random matrix theory \cite{b18}, $1/[(\hat{\mathbf{H}}^{H}\hat{\mathbf{H}})^{-1}]_{kk}$ can be rewritten as $\hat{\mathbf{h}}_{k}^{H}\tilde{\mathbf{M}}\hat{\mathbf{h}}_{k}$, where $\tilde{\mathbf{M}}$ is a non-negative Hermitian matrix with $K-e+f-1$ eigenvalues equal zero and $M-K+e-f+1$ eigenvalues equal 1. Since $\hat{\mathbf{h}}_{k} \sim \mathcal{CN}(0,\frac{\beta^2}{\beta+\tau_{\infty}^2}\mathbf{I})$, $1/[(\hat{\mathbf{H}}^{H}\hat{\mathbf{H}})^{-1}]_{kk}$ is a random variable with the PDF given as $g(x)= \frac{x^{\theta_{2}-1}e^{-x/2(\beta^2/(\beta+\tau_{\infty}^2))}}{2^{\theta_2}\Gamma(\theta_2)}$ ($x\geq 0$). The PDF of $\gamma_{k|ef}$ can be easily derived from $g\left(x\right)$ since it is a scaled version of $\frac{1}{[\left(\hat{\mathbf{H}}^{H}\hat{\mathbf{H}}\right)^{-1}]_{kk}}$, which is also Chi-square distributed.
\end{proof}

With the post-processing SNR distribution obtained in Lemma 3, the BLER of an active user that is successfully detected, is obtained in the following theorem.

\textbf{Theorem 1}.
    The BLER of an active user $k$ that is correctly detected, i.e., $k\in\mathcal{K}\cap \hat{\mathcal{K}}$, is given as follows:
\begin{align}
\begin{aligned}
\bar{\varepsilon}_{k}&=\sum\limits_{f=0}^{N-K}{N-K\choose f} P_{F}^{f} (1-P_{F})^{N-K-f}\\
&\times\left(\sum\limits_{e=0}^{K} {K\choose e} P_{M}^{e}(1-P_{M})^{K-e}\bar{\varepsilon}_{k|ef}\right),
\end{aligned}
\end{align}

\noindent where 
\begin{align}
    \bar{\varepsilon}_{k|ef}\approx\left\{
    \begin{aligned}
    &\mathbb{E}_{\gamma_{k|ef}}\left[\varepsilon_{k}\left(\gamma_{k|ef}\right)\right], M\geq|\hat{\mathcal{K}}|,\\
    & 1,M<|\hat{\mathcal{K}}|.
    \end{aligned}
    \right.
\end{align}

\begin{proof} For $M\geq|\hat{\mathcal{K}}|$, conditioned on an outcome of user activity detection with $e$ missed detection and $f$ false alarm users, the average BLER of an active user that is detected correctly can be derived by taking expectation on both sides of (11) with respect to $\gamma_{k|ef}$. For $M<|\hat{\mathcal{K}}|$, $\bar{\varepsilon}_{k|ef}$ is approximated as 1 due to the assumption on the ZF equalizer as imposed in Section II. By further incorporating the probabilities of different user activity detection outcomes, (14) can be obtained.
\end{proof}

Nonetheless, the BLER expression in (14) involves expectations on the Q-function, which are costly for numerical evaluation and prevent further insights. Next, we develop a tight approximation for (15) to simplify our analysis.

\textbf{Corollary 1}. Suppose $M\geq|\hat{\mathcal{K}}|$, the average BLER conditioned on $e$ missed detection and $f$ false alarm users in (16) is tightly approximated as follows:
\begin{align}
\bar{\varepsilon}_{k|ef}=\frac{\underline{\gamma}(\theta_2,\frac{2^{\frac{c}{d}}-1}{2\theta_1})}{\Gamma(\theta_2)},
\end{align}

\noindent where $\underline{\gamma}(\cdot)$ denotes the lower incomplete Gamma function.

\begin{proof} The proof is obtained by invoking a tight approximation of the BLER for finite-blocklength channel coding given as follows \cite{b19}:
\begin{align}
Q\left(\frac{C(\gamma)-R}{\sqrt{V(\gamma)/d}}\right) \approx\left\{\begin{array}{cc}
1, & \gamma \leq v \\
A\left(\gamma\right), & v <\gamma<\mu \\
0, & \gamma \geq \mu
\end{array}\right.,
\end{align}

\noindent where $A\left(\gamma\right)\triangleq 1/2-\chi\sqrt{d}(\gamma-r)$, $\chi\triangleq \sqrt{\frac{1}{2\pi (2^{\frac{2c}{d}}-1)}}$, $v\triangleq r-\frac{1}{2\chi\sqrt{d}}$, $\mu\triangleq r+\frac{1}{2\chi\sqrt{d}}$, and $r\triangleq 2^{R}-1=2^{\frac{c}{d}}-1$. By substituting the right-hand side of (17) into (15) with $\gamma = \gamma_{k|ef}$, $\bar{\varepsilon}_{k|ef}$ can be approximated as follows:
\begin{align}
\bar{\varepsilon}_{k|ef} \approx \chi \sqrt{d} \int_{v}^{\mu} F_{\gamma_{k|ef}}(x) dx,
\end{align}

\noindent where $F_{\gamma_{k|ef}}(x)\triangleq \frac{\underline{\gamma}(\theta_2,\frac{x}{2\theta_1})}{\Gamma(\theta_2)}$ is the cumulative distribution function (CDF) of the post-processing SNR. Then, by applying the ﬁrst-order Riemann integral approximation, (18) can be simplified as follows:
\begin{align}
\bar{\varepsilon}_{k|ef} \approx \chi \sqrt{d} (\mu-v) F_{\gamma_{k|ef}}\left(\frac{\mu+v}{2}\right)= \frac{\underline{\gamma}(\theta_2,\frac{2^{\frac{c}{d}}-1}{2\theta_1})}{\Gamma(\theta_2)}.
\end{align}

\end{proof}

Due to the identical user active probabilities and statistical channel inversion power control, the average BLER of the grant-free massive RA system is obtained as follows:
\begin{align}
    P_{e}=P_{M}+(1-P_{M})\bar{\varepsilon},
\end{align}

\noindent where $\overline{\varepsilon} = \overline{\varepsilon}_{k}, k\in \mathcal{N}$ can be evaluated via numerical integral with (15) or approximately in closed-form with (16).

\section{Pilot Length Optimization}
To the best of our knowledge, the above analysis is the first result that characterizes the data transmission performance of grant-free massive RA systems, accounting for the effects of activity detection and channel estimation errors, as well as considering short blocklength coding. This BLER analysis is valuable to provide practical system design guidelines, and this section presents a case study on optimizing the pilot length.

Given a fixed blocklength $T$ to transmit $c$ bits of payload data, the pilot length will affect activity detection, channel estimation, and data decoding in different ways. Specifically, increasing the pilot length will improve the performance of activity detection and channel estimation, but the corresponding channel coding rate will increase as well, which will weaken the error correction capability of the channel code. Thus, it can be expected that there exists an optimal pilot length which can balance the effects between activity detection/channel estimation and data decoding.

Our goal is to minimize the average BLER of the grant-free massive RA system given the number of payload bits and the transmission blocklength, as formulated in the following optimization problem:
\begin{align}
    \min _{L \in\{K+1, \cdots, T-c\}} P_{M}\left(L\right)+\left(1-P_{M}\left(L\right)\right)\bar{\varepsilon}(L).
\end{align}

\noindent In principle, we can evaluate the BLER for all possible $L$ in set $\{K+1, \cdots, T-c\}$ to optimally solve Problem (21). However, since the fixed point $\tau_{\infty}$ can only be obtained numerically by executing the AMP algorithm, enumerating all possible $L$ requires significant computation. To develop a computation-efficient pilot length optimization method, we consider the high SNR approximation of the state evolution that yields a simple analytical characterization of the ﬁxed point \cite{b12}, i.e., $ \lim_{\frac{\beta}{\sigma^{2}}\rightarrow \infty} \tau^{2}_{\infty} =  \frac{\sigma^{2}}{L-K}$. Also, as $N$ is large, the polynomial form of the BLER expression in (14) makes it difficult for further analysis. To resolve this issue, we propose to retain only the dominant term in (14) corresponding to the case with $e=f=0$, i.e., with no user activity detection error, considering the requirements on missed detection and false alarm probabilities are typically smaller than $0.1\%$ [20, Section 8.4]. Thus, with Corollary 1, (14) is simplified as follows:
\begin{align}
\bar{\varepsilon}_{k}\approx (1-P_{F})^{N-K}(1-P_{M})^{K}\bar{\varepsilon}_{k|00},
\end{align}

\noindent where $\bar{\varepsilon}_{k|00}=\frac{\underline{\gamma}(M-K+1,\frac{2^{\frac{c}{d}}-1}{2\theta_1})}{\Gamma(M-K+1)}$ and $\theta_1=\frac{\beta(L-K)}{\sigma^2(L+\sigma^2/\beta)}$. We will validate the rationality of (22) numerically in Section V.

\section{Simulation Results}
In simulations, we consider a single-cell uplink cellular network with 2000 users, which are uniformly distributed within a circular ring with the inner and outer radius as 0.05 and 1km, respectively. Key simulation parameters are listed in TABLE \ref{table1}.

\begin{table}[ht]
\caption{Simulation Parameters}
\centering
    \begin{tabular}{c|c|c|c}
    \hline
     Parameters & Values & Parameters & Values\\
    \hline
    $M$ & 100 & $T$ & 250\\
    $c$ & 50 & $\boldsymbol{\alpha}_{n}$ & $\mathcal{C} \mathcal{N}(\mathbf{0}, \mathbf{I}_{M})$\\
    $\beta$ & -123.8 dB & $\sigma^2$ & -109 dBm\\
    \hline
    \end{tabular}
\label{table1}
\end{table}

\subsection{Verification of the BLER Expressions}
\begin{figure}[t]
\centering
\includegraphics[width=2.9in]{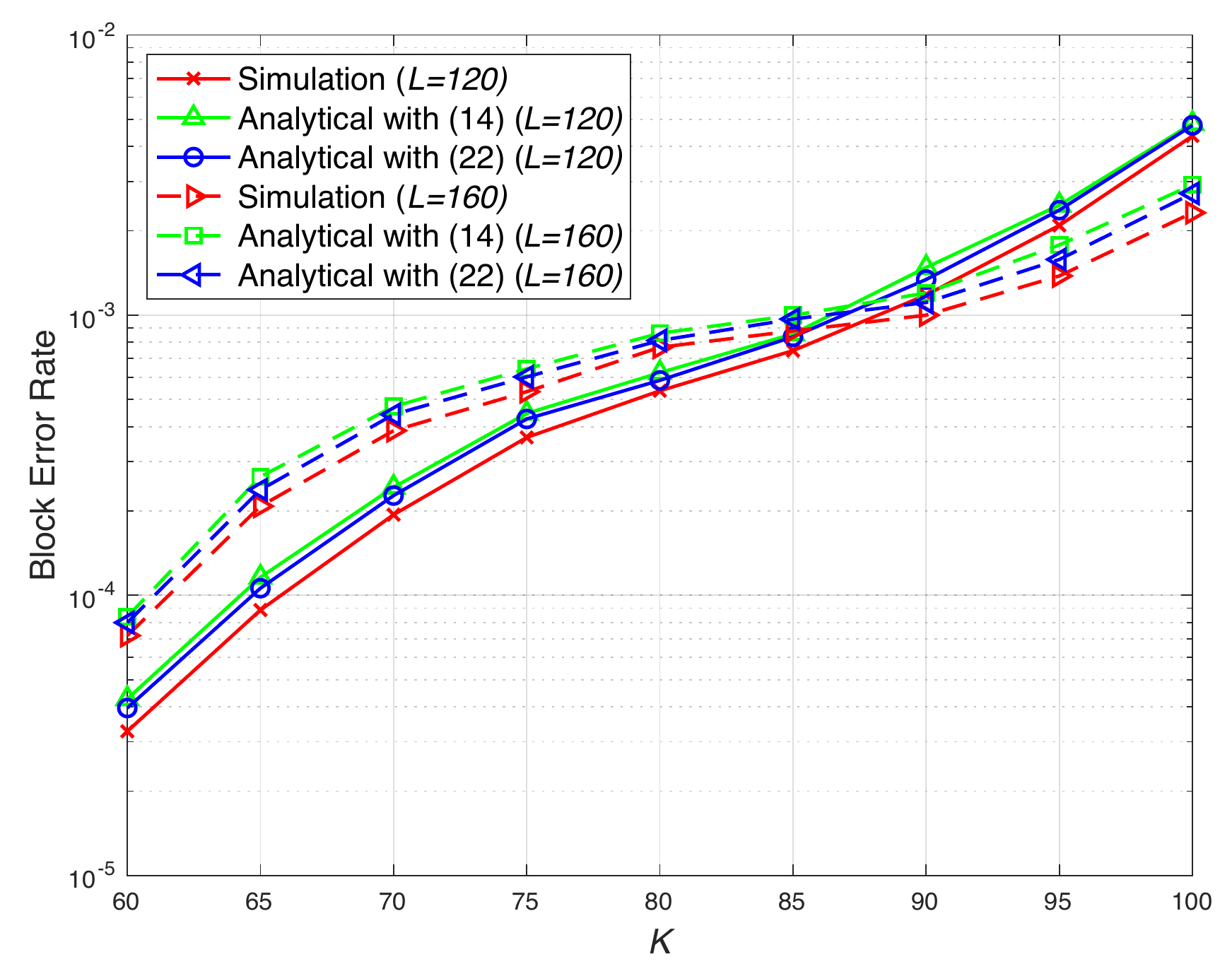}
\caption{BLER versus the number of active users ($\lambda = K \slash N$).}
\label{bler}
\end{figure}

We first verify the BLER expressions obtained respectively with $\bar{\varepsilon}_{k}$ in (14) (in which, $\{\varepsilon_{k|ef}\}$'s are computed via (15)) and (22) against the empirical results obtained via Monte-Carlo simulations. The empirical BLER values are averaged over $5000$ independent channel realizations. We examine two configurations with $L=120$ and $160$ pilot symbols in each transmission block, respectively, and show the relationship between the BLER and the number of active users in Fig. \ref{bler}. It is seen that the analytical curves match well with the empirical results in both pilot length configurations. We also observe that the BLER curves obtained with $\bar{\varepsilon}_{k}$ in (22) tightly lower bounds those obtained with $\bar{\varepsilon}_{k}$ in (14), which indicates that retaining only the dominant term is sufficient for accurate BLER analysis. It is noteworthy that the two sets of curves intersect at around $K = 87$. Specifically, when $K<87$, the BLER performance with $L=160$ is worse than that with $L=120$; and when $K\geq 87$, $L = 120$ turns out to be a better option. This observation validates the necessity of pilot length optimization for grant-free massive RA systems.
\addtolength{\topmargin}{0.02in}

\subsection{Pilot Length Optimization}
\begin{figure}[t]
\centering
\includegraphics[width=2.9in]{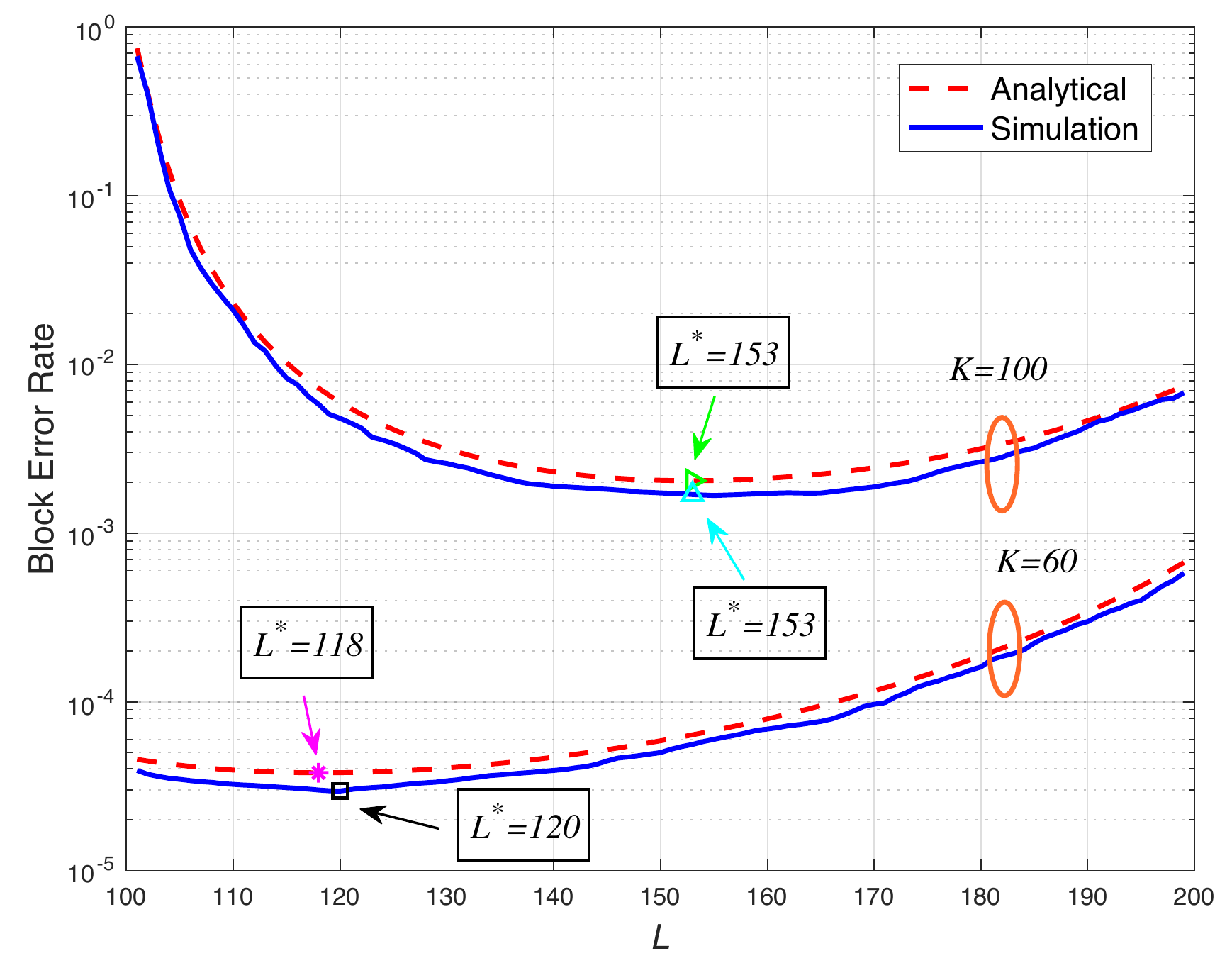}
\caption{BLER versus pilot length ($\lambda = K \slash N$).}
\label{optl}
\end{figure}

Next, we investigate the pilot length optimization in order to minimize the BLER. The relationship between the BLER and pilot length with $K = 60$ and $100$ is depicted in Fig. \ref{optl}. The analytical curves are generated based on the approximated expression of $\bar{\varepsilon}_k$ in (22), and the simulation results are obtained via 5000 independent channel realizations. We sweep the pilot length from $L=100$ to $200$, and accordingly, the channel coding rate increases from $1/3$ to $1$. From this figure, we again corroborate the theoretical analysis. Also, the BLER curves appear to be convex, which, in a sense confirms there is the best tradeoff between the pilot length and channel coding rate.

In the case with 100 active users, we see that when the pilot length is less than $150$, the BLERs decrease with $L$. In this regime, the system performance is limited by the accuracy of user activity detection and channel estimation, and thus adding a few more pilot symbols can reduce the BLER significantly. When the pilot length is greater than $150$, reserving more symbols for pilot transmission contributes adversely. This is because the plateau of the activity detection and channel estimation performance as well as the increase of the channel coding rate. From the simulation curve, the optimal pilot length is given by 153, which coincides with that obtained via the analytical curve. Similar behavior is observed for the case with $K=60$, where the optimal pilot length is 120 according to the simulation results, which is very close to 118 as obtained from the analytical curve. These results demonstrate the benefits of our analysis for efficient pilot length optimization.
\addtolength{\topmargin}{0.04in}

\section{Conclusions}
In this paper, we investigated the short-packet transmission performance with grant-free massive random access. Based on a two-phase transmission scheme, we adopted the AMP algorithm for joint activity detection and channel estimation. Then, with the results of AMP, we considered the block error rate (BLER) in the finite blocklength regime as the performance metric, and derived the post-processing SNR distribution by the random matrix theory in order to characterize the data transmission error. We further approximated our analytical result to a close-form expression, and then by applying our simple analytical results, the optimization of the pilot length was considered to make a trade-off between the pilot length and channel coding rate so that the minimum BLER can be achieved. Simulation results verified our theoretical analysis. Both the analysis and the corresponding optimization are promising to provide important design guidelines for grant-free massive random access systems, which call for more investigations.

\end{document}